\documentstyle[sprocl,epsfig]{article}

\newcommand{\modt}{\mbox{$|t|$}}

\newcommand{\qsq}{\mbox{$Q^2$}}

\newcommand{\rh}{\mbox{$\rho$}}

\newcommand{\ph}{\mbox{$\phi$}}

\newcommand{\jpsi}{\mbox{$J/\psi$}}

\newcommand{\rzzzz}{\mbox{$r_{00}^{04}$}}
\newcommand{\rzqzz}{\mbox{$r_{00}^{04}$}}
\newcommand{\rzquz}{\mbox{$r_{10}^{04}$}}
\newcommand{\rzqumu}{\mbox{$r_{1-1}^{04}$}}
\newcommand{\ruuu}{\mbox{$r_{11}^{1}$}}
\newcommand{\ruzz}{\mbox{$r_{00}^{1}$}}
\newcommand{\ruuz}{\mbox{$r_{10}^{1}$}}
\newcommand{\ruumu}{\mbox{$r_{1-1}^{1}$}}
\newcommand{\rduz}{\mbox{$r_{10}^{2}$}}
\newcommand{\rdumu}{\mbox{$r_{1-1}^{2}$}}
\newcommand{\rcuu}{\mbox{$r_{11}^{5}$}}
\newcommand{\rczz}{\mbox{$r_{00}^{5}$}}
\newcommand{\rcuz}{\mbox{$r_{10}^{5}$}}
\newcommand{\rcumu}{\mbox{$r_{1-1}^{5}$}}
\newcommand{\rsuz}{\mbox{$r_{10}^{6}$}}
\newcommand{\rsumu}{\mbox{$r_{1-1}^{6}$}}

\newcommand{\gev}{\mbox{\rm GeV}}

\newcommand{\gevsq}{\mbox{${\rm GeV}^2$}}

\newcommand{\gevsqm}{\mbox{${\rm GeV}^{-2}$}}

\newcommand{\bce}{\begin{center}}
\newcommand{\ece}{\end{center}}
\newcommand{\beq}{\begin{equation}}
\newcommand{\eeq}{\end{equation}}
\newcommand{\bea}{\begin{eqnarray}}
\newcommand{\eea}{\end{eqnarray}}
\def\lsim{\mathrel{\rlap{\lower4pt\hbox{\hskip1pt$\sim$}}
    \raise1pt\hbox{$<$}}}         
\def\gsim{\mathrel{\rlap{\lower4pt\hbox{\hskip1pt$\sim$}}
    \raise1pt\hbox{$>$}}}         
\newcommand{\pom}{{I\!\!P}}

\def\ar#1#2#3   {{\em Ann. Rev. Nucl. Part. Sci.} {\bf#1} (#2) #3}
\def\epj#1#2#3  {{\em Eur. Phys. J.} {\bf#1} (#2) #3}
\def\err#1#2#3  {{\it Erratum} {\bf#1} (#2) #3}
\def\ib#1#2#3   {{\it ibid.} {\bf#1} (#2) #3}
\def\ijmp#1#2#3 {{\em Int. J. Mod. Phys.} {\bf#1} (#2) #3}
\def\jetp#1#2#3 {{\em JETP Lett.} {\bf#1} (#2) #3}
\def\mpl#1#2#3  {{\em Mod. Phys. Lett.} {\bf#1} (#2) #3}
\def\nim#1#2#3  {{\em Nucl. Instr. Meth.} {\bf#1} (#2) #3}
\def\nc#1#2#3   {{\em Nuovo Cim.} {\bf#1} (#2) #3}
\def\np#1#2#3   {{\em Nucl. Phys.} {\bf#1} (#2) #3}
\def\pl#1#2#3   {{\em Phys. Lett.} {\bf#1} (#2) #3}
\def\prep#1#2#3 {{\em Phys. Rep.} {\bf#1} (#2) #3}
\def\prev#1#2#3 {{\em Phys. Rev.} {\bf#1} (#2) #3}
\def\prl#1#2#3  {{\em Phys. Rev. Lett.} {\bf#1} (#2) #3}
\def\ptp#1#2#3  {{\em Prog. Th. Phys.} {\bf#1} (#2) #3}
\def\rmp#1#2#3  {{\em Rev. Mod. Phys.} {\bf#1} (#2) #3}
\def\rpp#1#2#3  {{\em Rep. Prog. Phys.} {\bf#1} (#2) #3}
\def\sjnp#1#2#3 {{\em Sov. J. Nucl. Phys.} {\bf#1} (#2) #3}
\def\spj#1#2#3  {{\em Sov. Phys. JEPT} {\bf#1} (#2) #3}
\def\zp#1#2#3   {{\em Zeit. Phys.} {\bf#1} (#2) #3}
\newcommand{\schc}{\mbox{{\rm SCHC}}}

\begin{document}

\title{Hard Diffraction in Vector Meson Production at HERA}

\author{P. MARAGE}

\address{Universit\'e Libre de Bruxelles - CP 230, B-1050 Brussels, Belgium
\\E-mail: pmarage@ulb.ac.be}

\maketitle\abstracts{A review is presented of diffractive vector meson 
production 
at HERA, with stress on the investigation of ``hard'' features and comparisons
with theoretical predictions based on perturbative QCD approaches.\footnote
{Paper presented at the XXIIIth International Symposium on Multiparticle
Dynamics, Delphi, Greece, September 1998.}}
\section{Introduction}
%
Since the first data taking in 1992, the HERA high energy $e p$ collider 
has shown to be a powerful tool for the study of the strong interaction, in 
particular to test 
the domain of applicability and the relevance of several approximations of 
perturbative QCD (pQCD) in the field of diffraction.
\subsection {Total Cross Section and Diffraction}
%
Two major experimental discoveries were made at HERA for the understanding
of strong interactions and of hadron structure. 

First, the observation that, in the deep inelastic scattering (DIS) domain, 
the $\gamma^* p$ cross section increases rapidly with energy. 
This is attributed to an enhancement of the number of gluons in the proton, 
the gluon structure function 
$x \cdot G(Q^2,x)$ thus growing fast as $x$ decreases ($x$ is the Bjorken 
scaling variable: $x = Q^2 / 2 p \cdot q$, where $p$ and $q$ are, respectively, 
the proton and the intermediate photon four-momenta and $Q^2 = - q^2$; 
the $\gamma^* p$ centre of mass energy $W$ is given by $W^2 = Q^2 / x - Q^2$).

This ``hard'' behaviour differs from that of the total and the elastic 
hadron-hadron cross sections (closely related through the optical theorem), 
which are characterised by a ``soft'' energy dependence.
In the framework of Regge theory~\cite{Collins}, elastic scattering
is attributed at high energy to the exchange between the incoming hadrons of 
a colourless object, the pomeron $\pom$.
The energy dependence of the total cross section is proportional to 
$W^{2[1-\alpha_{\pom }(t)]}$, where the pomeron trajectory $\alpha_{\pom}(t)$
is parameterised as~\cite{DoLa,cudell_fit} 
\begin{equation}
\alpha_{\pom}(t) = \alpha_{\pom}(0) + \alpha^\prime \cdot t
                 \simeq 1.08 + 0.25 \ t ,
						\label{eq:soft_pom}
\end{equation}
$t$ being the square of the four-momentum transfer.

The second major discovery in DIS at HERA is the substantial contribution 
($8 - 10 \% $) 
of events formed of two hadronic subsystems separated by a large gap in 
rapidity, devoid of hadronic activity~\cite{Derrick}.

This process is similar to diffractive scattering in hadron-hadron 
interactions, where the incoming hadrons are excited without colour 
exchange. 
Diffraction thus forms an extension of elastic scattering and is dominated 
at high energy by pomeron exchange with the ``soft'' behaviour of 
eq.~(\ref{eq:soft_pom}).
The interesting feature at HERA was to observe diffraction as a leading twist
process in DIS.
\subsection {``Soft'' Vector Meson Production}
%
An important case of diffractive scattering is that of vector meson (VM) 
production, in particular when the proton remains intact in the reaction: 
$e + p \rightarrow e + p + VM$.

%
%
\begin{figure}[htb] \unitlength 1.0 cm
\vspace{-0.6cm}
\begin{center}
   \epsfig{file=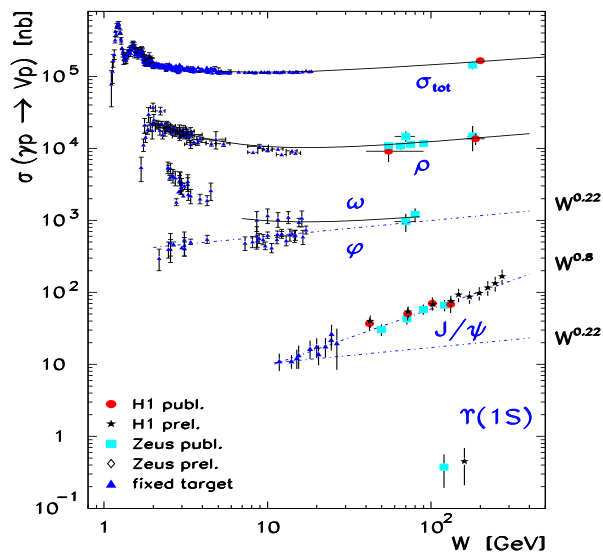,width=8.0cm,height=8.0cm}
\vspace{-0.5cm}
\caption{Total photoproduction and VM photoproduction cross sections, 
for fixed target and HERA experiments.
The solid curves are obtained from the ``soft'' pomeron 
parameterisation ~\protect\cite{DoLa}, with a decreasing contribution of 
reggeon exchange at low energy; the energy dependence 
noted $W^{0.22}$ is obtained from eq.~(\ref{eq:soft_pom}), taking into 
account the $t$ distribution of the events. 
The ``hard'', $W^{0.8}$, dependence is also shown in the case of \jpsi\ 
photoproduction.}
\label{fig:Wdep}
\vspace{-0.2cm}
\end{center}
\end{figure}
%
%
In the vector meson dominance (VDM) approach, the $J^{PC} = 1^{--}$ photon 
is modelled as the superposition of the lightest VM's ($\rho$, $\omega$, 
$\phi$).
The total $\gamma p$ cross section is thus expected to present the 
characteristic ``soft'' behaviour of hadron-hadron interactions. 
The production of light VM's, which is directly related to 
elastic scattering (with a differential absorption by the target proton of 
some of the hadronic components of the photon) is also expected to
present a ``soft'' energy dependence.
The gross features of this interpretation are supported by a huge quantity 
of data accumulated by fixed target experiments~\cite{bauer,CHIO,NMC,E665}.
At high energy, the HERA experiments have measured the total  
cross section in photoproduction 
($Q^2 \approx 0$)~\cite{z_sigmatot,h_sigmatot} and the cross section for 
diffractive photoproduction of 
$\rho$~\cite{zeus_phot,h_rho_phot,zeus_phott,zeus_phottt},
$\omega$~\cite{z_om_phot}, $\phi$~\cite{z_phi_phot}.
They exhibit the ``soft'' energy dependence described by  
parameterisation (\ref{eq:soft_pom}), as shown on Fig.~\ref{fig:Wdep}
(at low energy, a contribution from reggeon exchange, decreasing with $W$, is
present for $\sigma_{tot}$, \rh\ and $\omega$).
The W dependence of \rh\ photoproduction, studied as a function of $t$,
has also allowed measuring the slope $\alpha^\prime$ of the pomeron
trajectory~\cite{z_rh_alphaprim}.
%
%
\begin{center}
\begin{figure}[htb] \unitlength 1.0 cm
\vspace{-0.7cm}
\begin{picture}(12.0,8.0)
   \put(-0.1,0.7){\epsfig{file=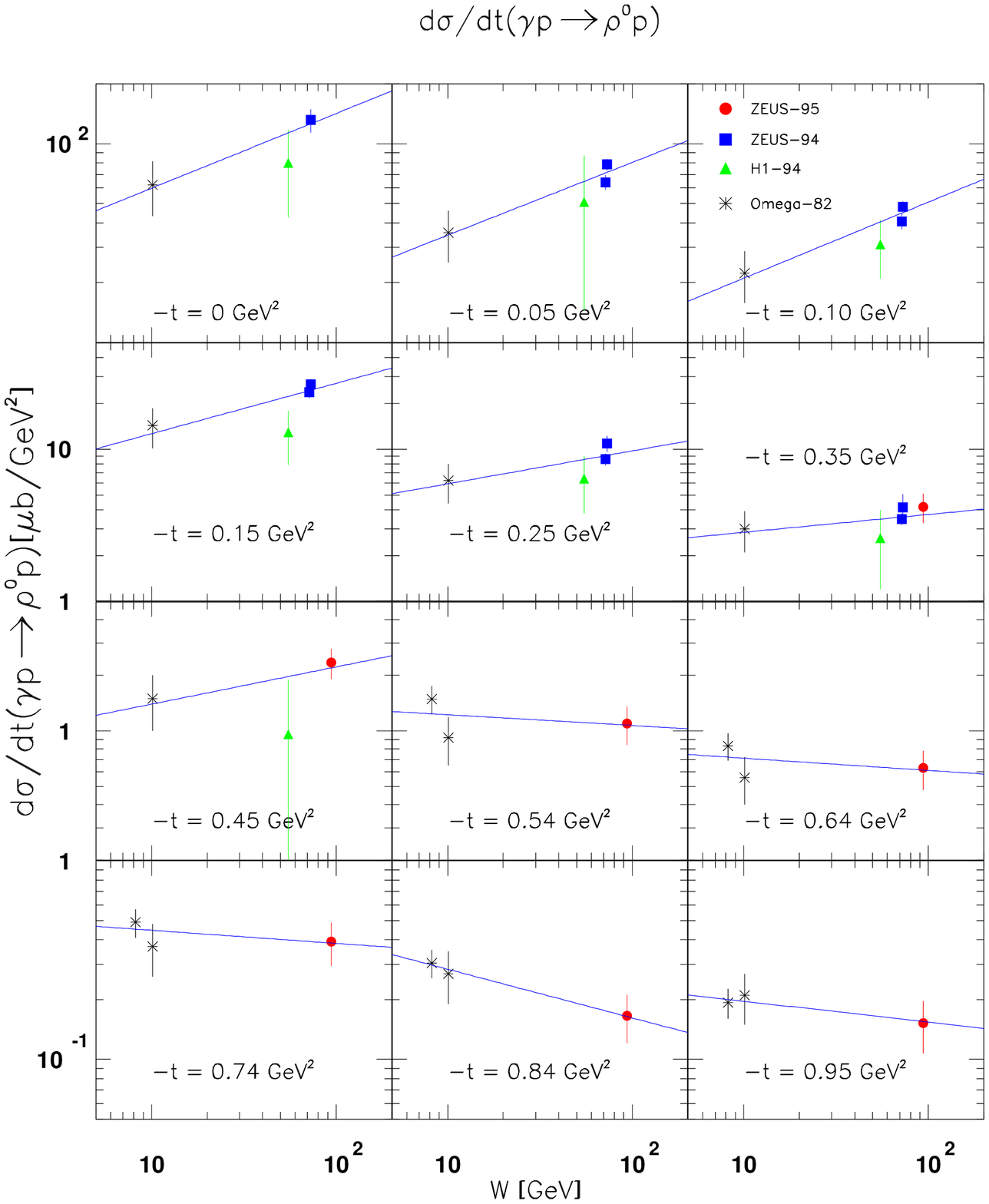,%
            bbllx=20pt,bblly=154pt,bburx=523pt,bbury=667pt,%
            width=6.0cm,height=6.0cm}}
   \put(6.5,1.1){\epsfig{file=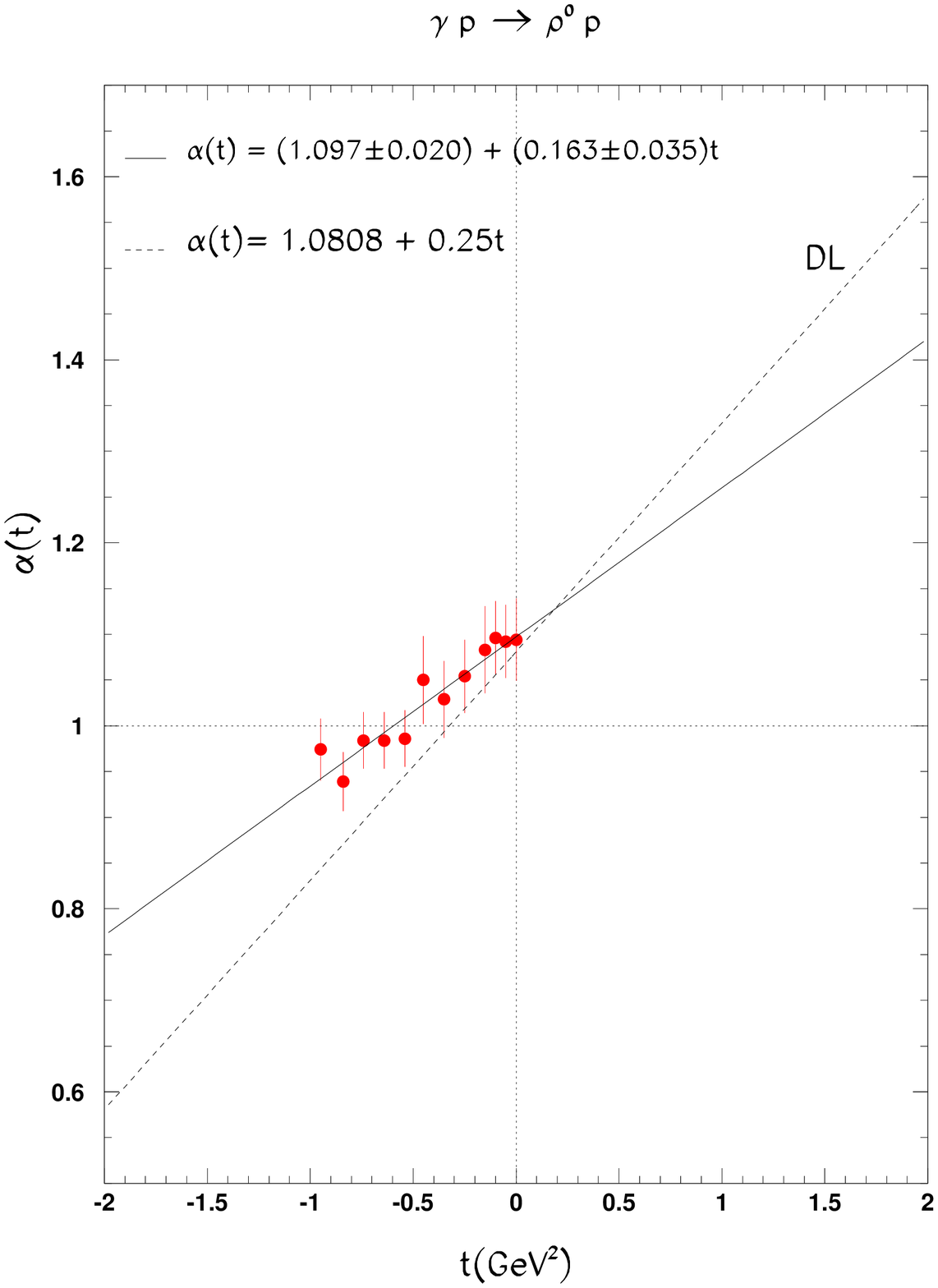,%
            bbllx=20pt,bblly=154pt,bburx=523pt,bbury=667pt,%
            width=5.2cm,height=4.8cm}}
\end{picture}
 \vspace{-0.8cm}
\caption{a) $W$ dependence of \rh\ meson photoproduction, in several bins in 
$t$~\protect\cite{z_rh_alphaprim}.
b) Slope of the pomeron trajectory for the reaction 
$\gamma p \rightarrow \rho p$ as obtained from a); the
dotted curve represents the parameterisation of~\protect\cite{DoLa}.}
\label{fig:alphaprim}
\vspace{-0.6cm}
\end{figure}
\end{center}
%
%
%
\subsection {``Hard'' Vector Meson Production and QCD}
%
At HERA, the main interest is for the production of light VM's at high \qsq\
or high \modt, and for the production of heavy vector mesons.
This is because two far-reaching questions can be raised:

1. Is the ``soft'', hadron-like behaviour observed in light VM 
photoproduction also observed in the presence of a ``hard'' scale: high \qsq,
high \modt\ or large quark mass ($c$, $b$) ?

2. In the presence of a ``hard'' scale, what are the relevant assumptions 
and approximations in pQCD calculations required to describe diffractive 
VM production ? Can this shed light on the partonic nature of the pomeron ?

A large number of experimental studies have thus been performed at HERA to 
investigate these questions.
Data have been collected, in the presence of the scales $Q^2$, $m_q$ and $t$, 
on the production of 
$\rho$, $\omega$, $\phi$, $\rho^\prime$, $J/\psi$, $\psi(2s)$ and $\Upsilon$ 
mesons, with studies of the differential $Q^2$, $W$ and $t$ distributions, of the 
polarisation characteristics, of the cross section ratio between several 
VM production and of the mass shape.
Only a small fraction of these results will be presented here.
They are largely based on results presented at the 
29th Int. Conf. on HEP held at Vancouver, Canada, in July, 1998.
%
%
\begin{center}
\begin{figure}[htb] \unitlength 1.0 cm
\vspace{-2.5cm}
\begin{picture}(10.0,4.5)
\put(2.7,-0.6){\epsfig{file=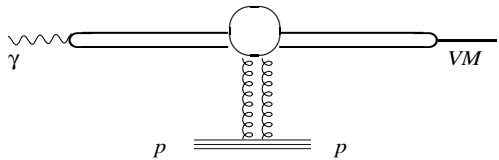,%
          bbllx=74pt,bblly=321pt,bburx=535pt,bbury=503pt,%
          height=2.4cm,width=6.7cm}}
\put(6.0,1.6){\oval(0.67,0.65)}
\end{picture}
 \vspace{-0.3cm}
\caption{Schematic representation of VM production in pQCD.}
\label{fig:factor}
\vspace{-0.5cm}
\end{figure}
\end{center}
%
%

A large number of theoretical papers based on pQCD has also
been published, presenting predictions for VM production under various 
assumptions and approximations 
(see e.g.~\cite{kopel,ryskin,Brodsky,fks,mrt,ivanov,royen}).
A general feature of these approaches is that, at high energy, the amplitude 
is factorised in three contributions, characterised by very different time
scales (see Fig.~\ref{fig:factor}):
\begin{equation}
A = \Psi^*_{\gamma^* \rightarrow q \bar q} 
    \otimes M_{q \bar q + p \rightarrow q \bar q + p} 
        \otimes \Psi_{q \bar q \rightarrow V}  .
					\label{eq:factor}
\end{equation}
The first factor corresponds to the amplitude for a long distance fluctuation 
of the photon into a $q \bar q$ pair.
The second factor describes the (short-time) scattering amplitude of this 
hadronic state with the proton. 
The exchange is generally modelled as a gluon pair (i.e. a colour singlet 
system), with $M \propto |x \cdot G(K^2,x)|^2$, the square of the gluon 
density in the proton. 
The order of magnitude of the scale $K^2$ at which the gluon structure 
function is probed is
%
$K^2 \simeq 1 / 4 \ (Q^2 + m_{V}^2 + \modt)$, 
%
since these three variables contribute to the ``resolution'' of the 
process; the factor $1 / 4$ 
comes from the sharing of the momenta between the two quarks.
The third factor in eq.~(\ref{eq:factor}) accounts for the recombination of 
the scattered hadronic state in the VM wave function.

However, as stressed e.g. in~\cite{fks}, theoretical calculations are
affected by significant uncertainties concerning the choice of the QCD scale,
of the gluon distribution and of the VM wave function, in particular the
effects of Fermi motion of the quarks within the meson.

\section{Differential Distributions}
%
\subsection{$W$ Dependence}
%
The most striking manifestation of pQCD features in VM production is  
to be expected in the $W$ dependence of the cross section, since the latter 
is related to the square of the gluon density in the proton, which increases 
rapidly with $W$ in the presence of a hard scale.
%
%
\begin{center}
\begin{figure}[htb] \unitlength 1.0 cm
\begin{picture}(12.0,6.0)
\vspace{-0.8cm}
   \put(0.0,0.0){\epsfig{file=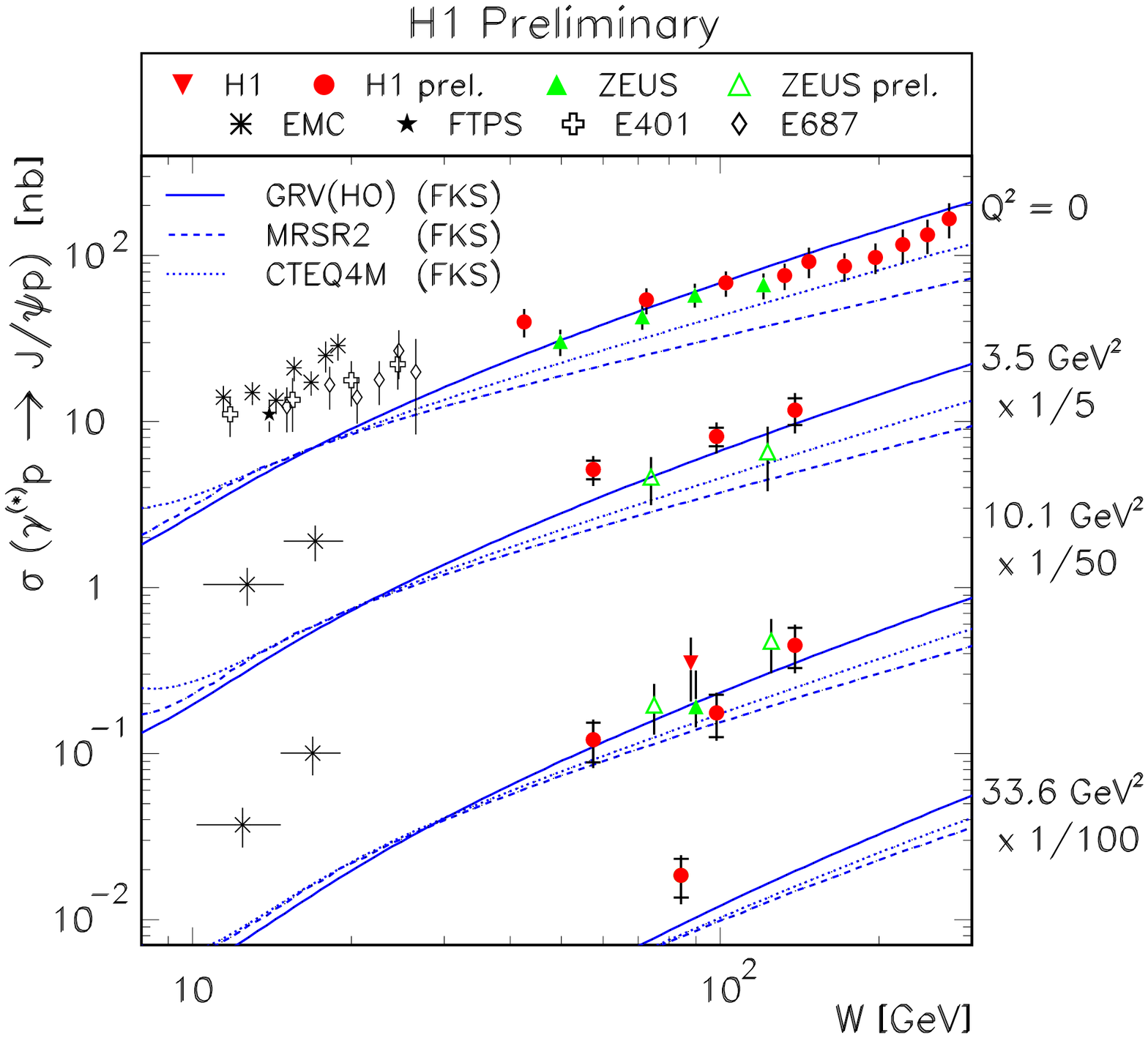,%
            width=6.0cm,height=6.0cm}}
   \put(6.5,0.0){\epsfig{file=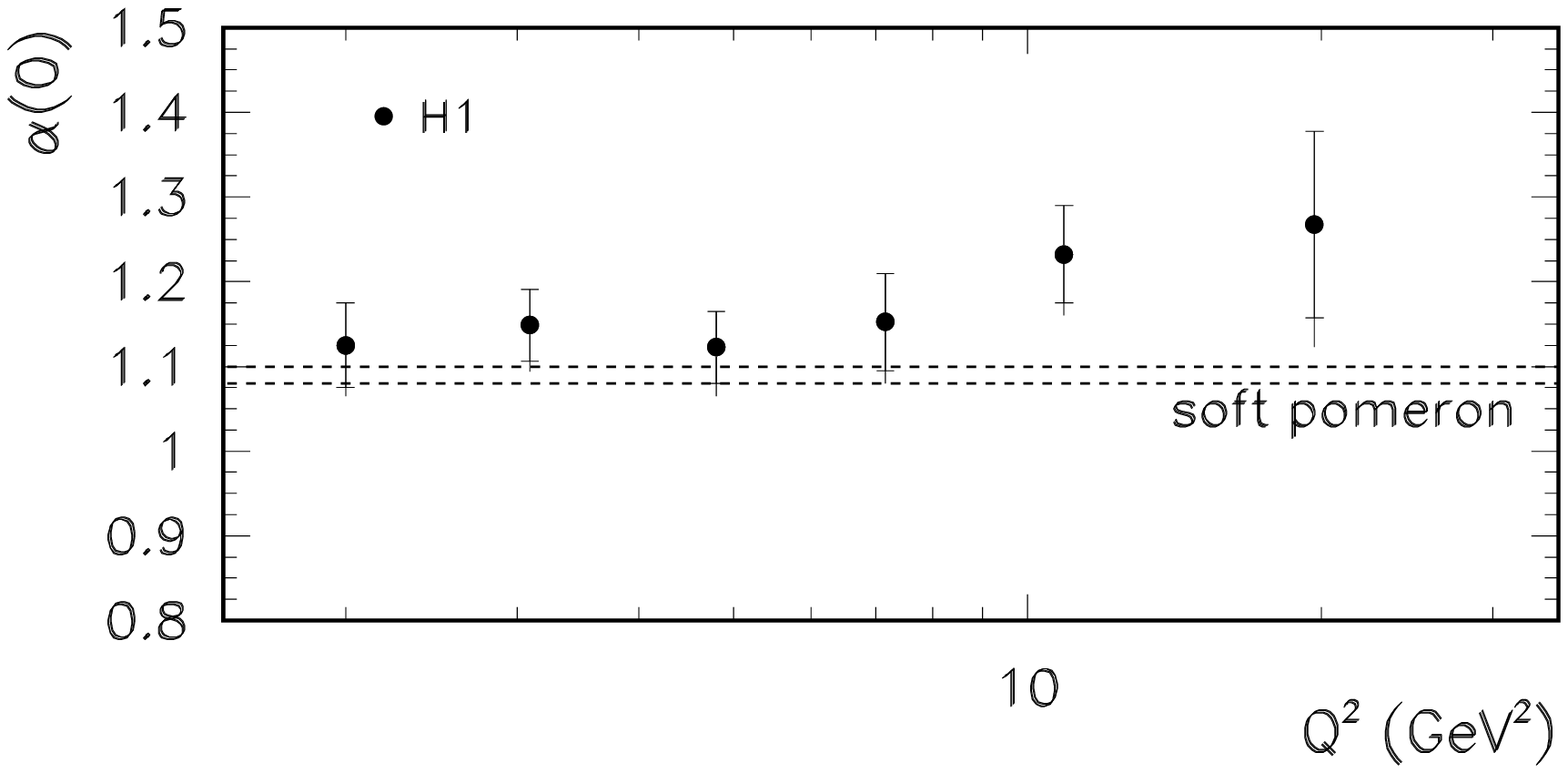,
            width=5.7cm,height=5.0cm}}
\end{picture}
\vspace{-0.8cm}
\caption{a) $W$ dependence of \jpsi\ meson photo- and electroproduction;
the curves represent predictions of a pQCD model~\protect\cite{fks} for
different gluon distribution functions in the proton and 
$m_c = 1.5$ \gev~\protect\cite{h_jpsi_DIS}.
b) \qsq\ dependence of the intercept $\alpha_\pom(0)$ for \rh\ meson 
electroproduction~\protect\cite{h_rho_DIS}; 
the dashed lines represent the range of values for the ``soft'' 
pomeron intercept, as derived from fits to the total and elastic 
hadron-hadron cross section measurements~\protect\cite{DoLa,cudell_fit}.}
\label{fig:Wdephard}
\end{figure}
\vspace{-0.5cm}
\end{center}
%
%
%

A ``hard'' behaviour is observed in photoproduction of \jpsi\ 
mesons~\cite{z_jpsi_phot,h_jpsi_phot,h_jpsi_photoprod}, as shown in 
Figs.~\ref{fig:Wdep} and~\ref{fig:Wdephard}a.
When the $W$ dependence of the cross section is parameterised as 
$\propto W^{\delta}$ (in a Regge approach, 
$\delta = 4 \alpha_\pom(\langle t \rangle)$), 
one finds $\delta \simeq 0.8$ for \jpsi\ photoproduction.
The contrast is thus manifest with the ``soft'' behaviour of light VM 
photoproduction, for which $\delta = 0.20 - 0.25$ (this value is in
agreement with the parameterisation of eq.(\ref{eq:soft_pom}), taking into
account the $t$ distribution). 
A similar behaviour is observed for \jpsi\ 
electroproduction~\cite{h_jpsi_DIS,z_rho_jpsi_DIS} 
(Fig.~\ref{fig:Wdephard}a).
The curves on this figure represent the predictions of a model
based on pQCD calculations~\cite{fks}, for different parameterisations of
the gluon distribution in the proton.
The agreement of these predictions with the data, especially as to the
shape of the distribution (the absolute values are
sensitive to the input charm quark mass), supports the modelisation of the 
pomeron as a colour-singlet gluon pair. 

For \rh\ and \ph\ meson electroproduction, the ``hard'' scale is related 
to \qsq.
Although the precision of the data is still limited, an indication is
present of a steeper $W$ dependence of the $\gamma^* p$ cross section 
as $Q^2$ increases for the \rh~\cite{z_rho_jpsi_DIS,h_rho_DIS}
and for the \ph~\cite{z_phi_DIS}.
This is shown for the \rh\ on Fig.~\ref{fig:Wdephard}b, where the pomeron 
intercept $\alpha_{\pom}(0)$ is plotted.

\subsection{$Q^2$ Dependence}
%
The cross section for \rh\ production in the DIS domain is presented 
as a function of \qsq\ on Fig.~\ref{fig:Q2_b}a for the 
ZEUS~\cite{z_rho_jpsi_DIS} and H1~\cite{h_rho_DIS} experiments, which
are in agreement.
The \qsq\ dependence is well parameterised in this domain as 
${\rm d} \sigma / {\rm d} Q^2 \propto 1 / {(Q^2+m_V^2)}^n$, 
with $n \simeq 2.28 \pm 0.06$ (combined value).
This behaviour is expected from pQCD calculations, which give for the 
(dominant - see below) longitudinal cross section~\cite{Brodsky}:
$\sigma_L \propto [\alpha_s(Q^2) \cdot x G(Q^2,x)]^2 / Q^6$, when
taking into account the $Q^2$ dependence in $\alpha_s(Q^2)$ and in $x
G(Q^2,x)$, as well as other uncertainties affecting the 
calculations~\cite{fks}.
Over the full measurement range, including photoproduction, the \qsq\ 
dependence of \rh\ cross section is best described by the QCD based
model of ref.~\cite{royen}.

For \ph\ production~\cite{z_phi_DIS}, a value similar to that for the \rh\ is
found.
For \jpsi\ production, the values $n = 2.24 \pm 0.19$~\cite{h_jpsi_DIS} and 
$n = 1.58 \pm 0.25$~\cite{z_rho_jpsi_DIS} are obtained.

\subsection{$t$ Dependence}
%
For not too large \modt\ values,
the $t$ 
distribution of VM production can reasonably well be parameterised in the 
exponential form ${\rm d} \sigma / {\rm d} t \propto e^{- b |t|}$.
In an optical model approach of diffraction, the slope parameter $b$ is
related to the convolution of the sizes of the interacting objects:
$b \propto R_p^2 + R_{q \bar q}^2$, with the proton radius $R_p$ giving a
contribution of the order of $4 - 5$~\gevsqm.

\begin{figure}[htbp] \unitlength 1.0 cm
\begin{center}
\vspace{-1.0cm}
\begin{picture}(12.0,6.0)
   \put(0.0,0.0){\epsfig{file=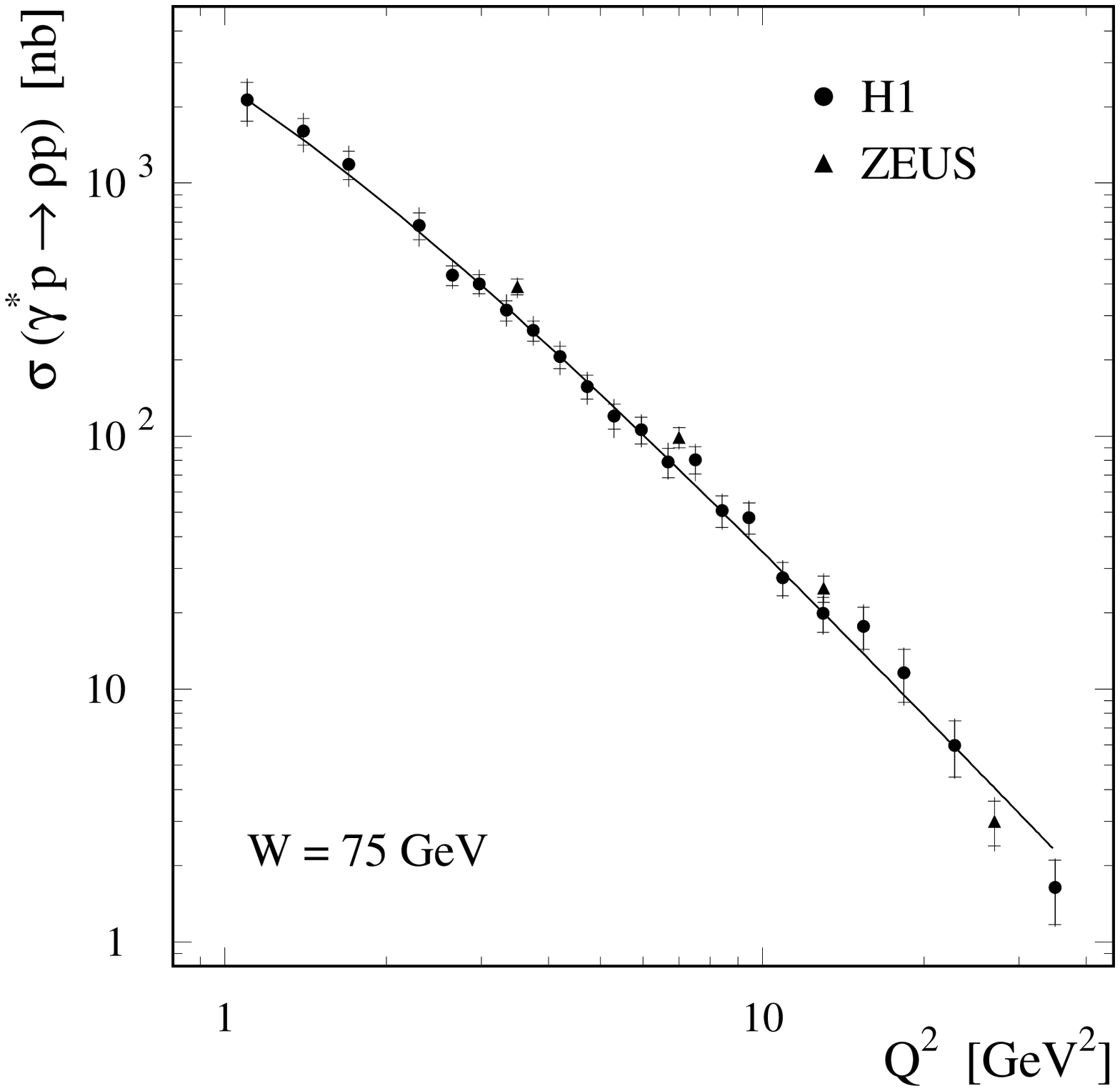,%
         height=5.8cm,width=5.8cm}}
   \put(6.0,0.1){\epsfig{file=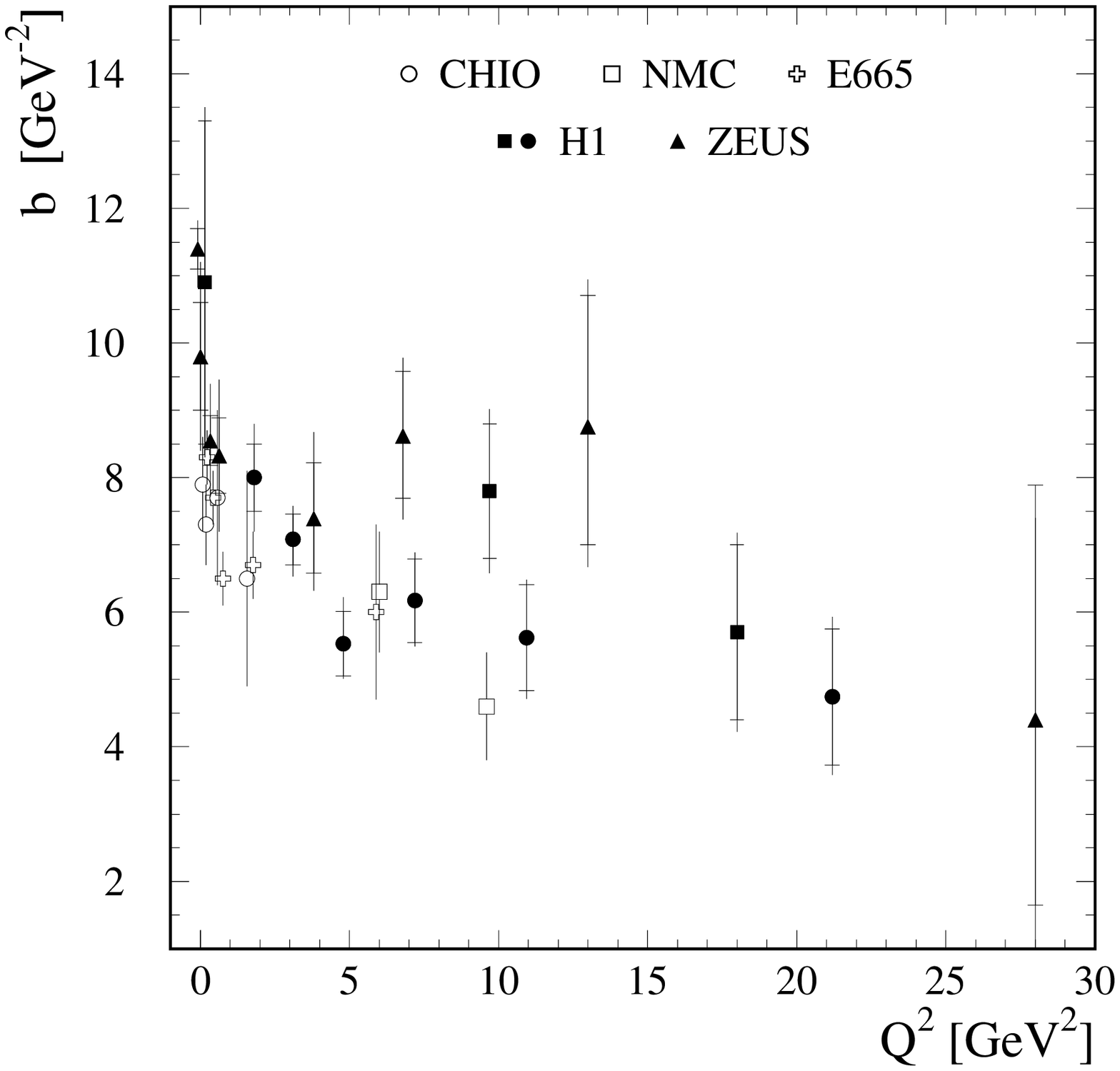,%
         height=5.5cm,width=5.8cm}}
\end{picture}
\end{center}
\vspace{-0.5cm}
\caption{a)~\qsq\ dependence of \rh\ production cross 
section~\protect\cite{z_rho_jpsi_DIS,h_jpsi_DIS}; the superimposed
curve is for $n = 2.24$.
b)~\qsq\ dependence of the slope parameter $b$ for elastic \rh\ production:
fixed target measurements of ref.~\protect\cite{CHIO,NMC,E665} and
HERA measurements of 
ref.~\protect\cite{z_rho_jpsi_DIS,H1_rho_94,h_rho_DIS} . }
\label{fig:Q2_b}
\vspace{-0.3cm}
\end{figure}

As observed in Fig.~\ref{fig:Q2_b}b, the slope $b$ for \rh\ 
production decreases when \qsq\ increases, in agreement with the decrease of 
the transverse size of the virtual $q \bar q$ pair expected in pQCD 
calculations. 

For \jpsi\ photo- and electroproduction, a slope of the order of 
$b \simeq 4-5$ \gevsqm\ is 
measured~\cite{h_jpsi_phot,z_jpsi_phot,z_rho_jpsi_DIS,h_jpsi_DIS}, 
confirming the small size 
of the \jpsi\ meson.

%

Fig.~\ref{fig:Q2_b}b also suggests that, at low \qsq, the slope parameter 
$b$ for \rh\ production increases from the fixed target to the HERA energy 
range. 
This behaviour, known as ``shrinkage'' and expected in Regge theory, is 
related to the non-zero slope $\alpha^\prime$ of the ``soft'' pomeron 
trajectory.
In contrast, no shrinkage is expected in a pQCD approach for asymptotically 
high values of the QCD scale ($\alpha^\prime \approx 0$).
However, no significant measurement has been possible so far using the HERA 
experiments only, neither for \rh\ production at high \qsq\ nor for \jpsi\ 
production, and the conclusions to be drawn from comparisons between
fixed target and HERA data remain 
controversial~\cite{z_rh_alphaprim,h_shrinkage_jpsi}.

\section{Polarisation}
%
The measurement of VM decay angular distributions allows the  
determination of the spin density matrix elements, which are related to the 
helicity amplitude $T_{\lambda_{V} \lambda_{\gamma}}$, where 
$\lambda_{V}$ and $\lambda_{\gamma}$ are the helicities of the VM
and of the photon, respectively~\cite{schilling-wolf}.
In the case of $s$-channel helicity conservation (\schc), the helicity of
the photon is retained by the VM and the matrix elements containing  
helicity changing amplitudes ($\lambda_{V} \neq \lambda_{\gamma}$) 
are thus zero.

Measurements of the full set of matrix elements have been performed
for \rh\ as a function of \qsq\ (Fig.~\ref{fig:matqsq}a), $W$ and 
$t$~\cite{h_rho_DIS,z_ang}, and for $\phi$ mesons~\cite{z_ang}.

As is visible on Fig.~\ref{fig:matqsq}, the data are compatible with \schc,
except for a small but significant deviation from zero of the  
matrix element \rczz.
The helicity flip amplitude $T_{\lambda_{\rho} \lambda_{\gamma}} = T_{01}$
is thus determined to be $8 \pm 3 \%$ of the non-flip amplitudes
$\sqrt {T_{00}^2 + T_{11}^2}$.
This value is of the order of magnitude of that found at lower energy and
lower \qsq~\cite{CHIO,joos}.

Neglecting the small violation of \schc\ (which would affect the value of $R$ 
by $2.5 \pm 1.5 \%$), the matrix element \rzqzz\
can be used to extract the ratio $R$ of cross sections for
\rh\ production by longitudinal and transverse virtual photons:
$R = \sigma_L / \sigma_T = \rzzzz / \varepsilon \cdot (1-\rzzzz)$,
where $\varepsilon$ is the polarisation parameter 
($\langle \varepsilon \rangle = 0.99$ at HERA).
Fig.~\ref{fig:matqsq}b shows that R rises steeply at small \qsq, and that
the longitudinal $\gamma^*p$ cross section dominates over the transverse 
cross section for \qsq\ $\gsim$ 2~\gevsq.
However, the rise is non-linear, with a weakening dependence at large \qsq\
values, and $R$ is $\approx$ 3 for \qsq\ $\gsim 10-20$~\gevsq.

This feature is not reproduced by numerous models based on VDM or QCD, which
predict a linear increase of $R$ with \qsq.
However, the model of ref.~\cite{royen}, based on QCD, gives a good 
description of $R$ over the full \qsq\ range, as does also a model based
on GVDM~\cite{sss}.
Another model based on QCD~\cite{mrt} predicts a moderate increase of $R$ 
with \qsq\ in the DIS domain.

\begin{center}
\vspace*{1.9cm}
\begin{figure}[htbp] \unitlength 1.0cm
\begin{picture}(0.0,0.0)
\put(-0.5,0.0){\epsfig{file=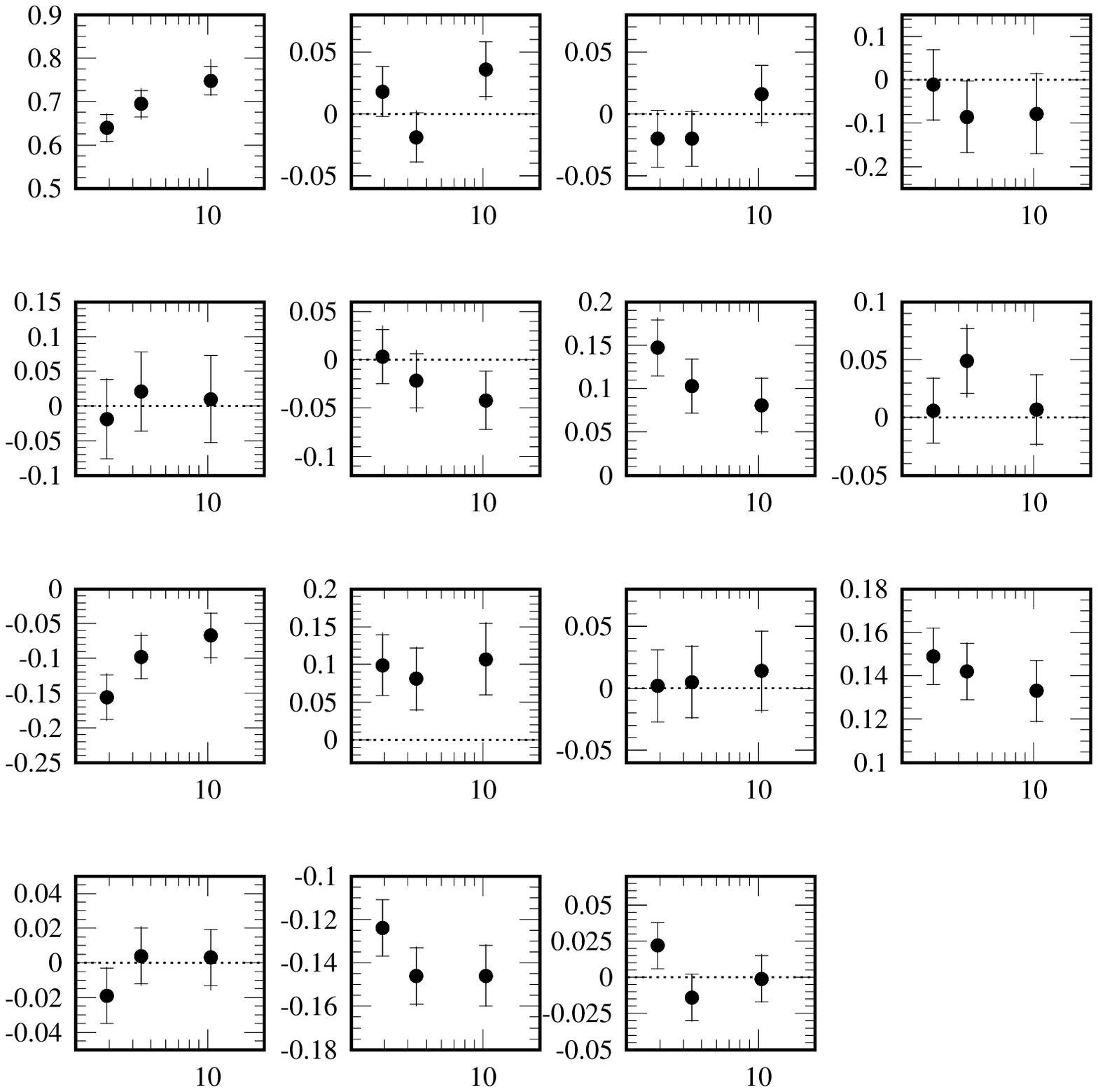,%
            bbllx=55pt,bblly=189pt,bburx=450pt,bbury=693pt,%
            height=8.5cm,width=6.5cm}}
\put(0.2,7.85){\large \rzqzz}
\put(2.1,7.85){\large {\rm Re} \rzquz}
\put(4.0,7.85){\large \rzqumu}
\put(5.9,7.85){\large \ruzz}
\put(0.2,5.82){\large \ruuu}
\put(2.1,5.82){\large {\rm Re} \ruuz}
\put(4.0,5.82){\large \ruumu}
\put(5.9,5.82){\large {\rm Im} \rduz}
\put(0.2,3.8){\large {\rm Im} \rdumu}
\put(2.4,3.8){\large \rczz}
\put(4.0,3.8){\large \rcuu}
\put(5.9,3.8){\large {\rm Re} \rcuz}
\put(0.2,1.8){\large \rcumu}
\put(2.1,1.8){\large {\rm Im} \rsuz}
\put(4.0,1.8){\large {\rm Im} \rsumu}
\put(5.7,1.2){\Large H1}
\put(5.7,0.5){\large \qsq~[\gevsq] }
\end{picture}
\put(7.3,1.0){\epsfig{file=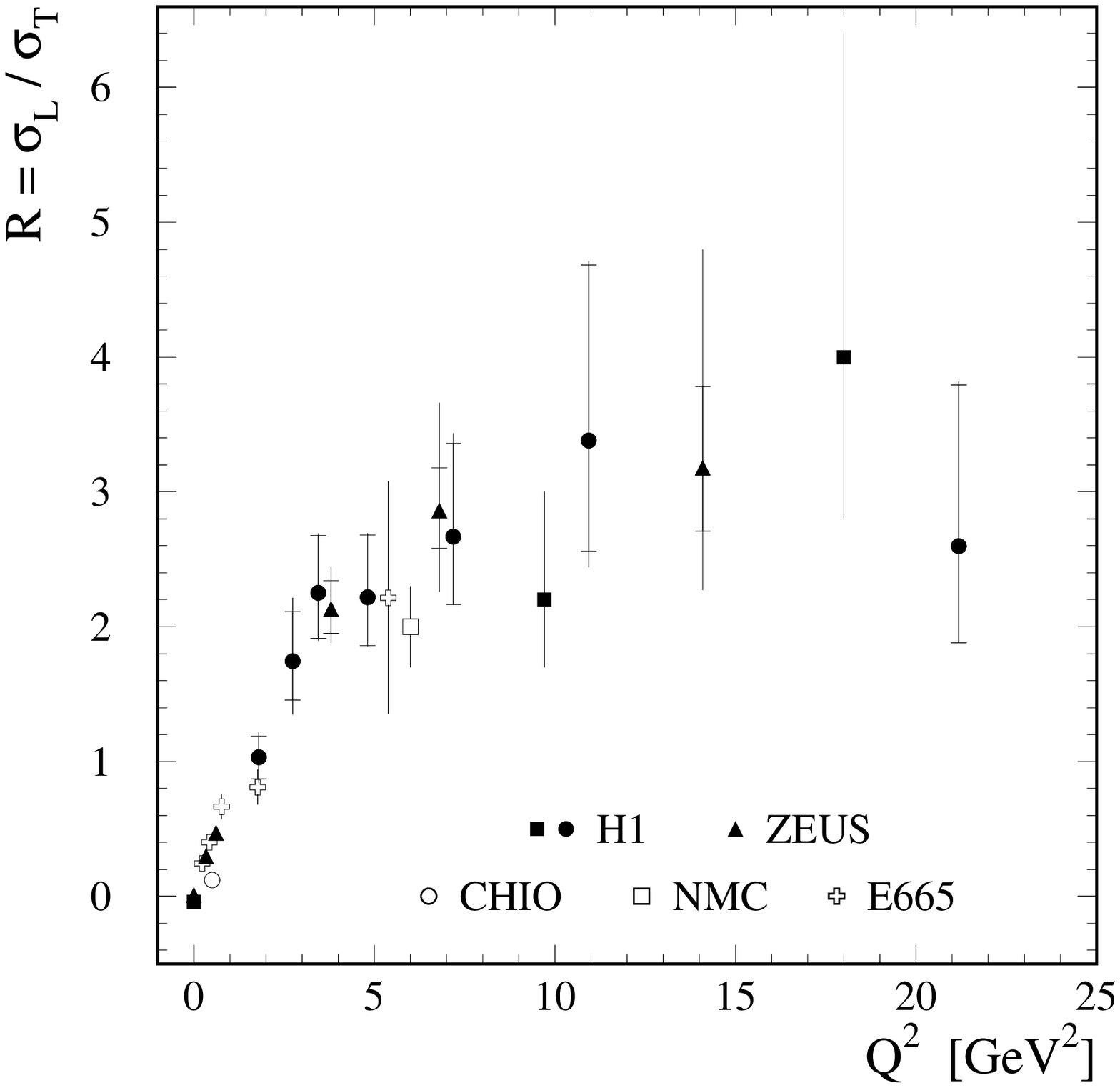,%
         height=5.0cm,width=5.cm}}
\vspace*{-0.4cm}
\caption{a) Spin density matrix elements for elastic electroproduction 
  of \rh\ mesons as a function of \qsq~\protect\cite{h_rho_DIS};   
  the dashed lines indicate the expected null values in the case of SCHC.
  b) The ratio $R$ of cross sections for elastic \rh\ meson electroproduction 
  by longitudinal and transverse photons,  as a function of \qsq.}
\label{fig:matqsq}
\end{figure}
\vspace*{-0.5cm}
\end{center}

It is also found~\cite{h_rho_DIS} that the longitudinal and transverse 
amplitudes are nearly in phase ($\cos \delta = 0.93 \pm 0.03$), assuming 
\schc\ and natural parity exchange. This is similar to lower energy 
measurements~\cite{CHIO,joos,delpapa}

A QCD based calculation~\cite{ivanov} predicts for the amplitudes the 
hierarchy 
\begin{equation}
|T_{\lambda_{\rho} \lambda_{\gamma}}| = 
                  |T_{00}| > |T_{11}| > |T_{01}| > |T_{10}| > |T_{1-1}|,
                                 \label{eq:hierarchy}
\end{equation}
which is supported by the measurement of the matrix elements, and also
the magnitude of the element \rczz~\cite{h_rho_DIS,z_ang}.

Values of the matrix elements close to those for the \rh\ are obtained
for \ph\ mesons~\cite{z_ang}.
For \jpsi, the ratio $R$ of cross sections increases from values compatible 
with zero in photoproduction~\cite{h_jpsi_phot,z_jpsi_phot}
to $\approx 0.4$ for $\langle Q^2\rangle \simeq 4$ 
\gevsq~\cite{z_rho_jpsi_DIS,h_jpsi_DIS}; this is smaller than
for \rh\ production at the same $\langle Q^2\rangle$, but is of the same 
order if compared at the same value of $Q^2 / m_V^2$.

\section{Other Features}
%
\subsection{VM Production Ratio}
%
Predictions are obtained in pQCD for the cross section ratio of different VM
production~\cite{kopel,fks}. 
As apparent in eq. (\ref{eq:factor}), this ratio is determined by the photon
coupling to the $q \bar q$ pairs, i.e. the charge of the quarks in the VM's,
and the effects of the wave functions.
For $\phi / \rho$~\cite{z_phi_phot,z_phi_DIS,H1_phi,H1_rho_95}, the 
ratio increases with \qsq\ towards the value $2 / 9$
obtained from quark counting (see Fig.~\ref{fig:VM_ratio}a).
For $\psi / \rho$, the ratio is about a factor $1/200$ in photoproduction 
in the HERA energy range, but flavour symmetry is restored within a factor 2
for \qsq\ above 10 \gevsq~\cite{z_rho_jpsi_DIS,H1_rho_94}.

\begin{center}
\begin{figure}[htbp] \unitlength 1.0 cm
 \vspace{-0.8cm}
  \begin{picture}(12.0,6.0)
   \put(-0.3,0.0){\epsfig{file=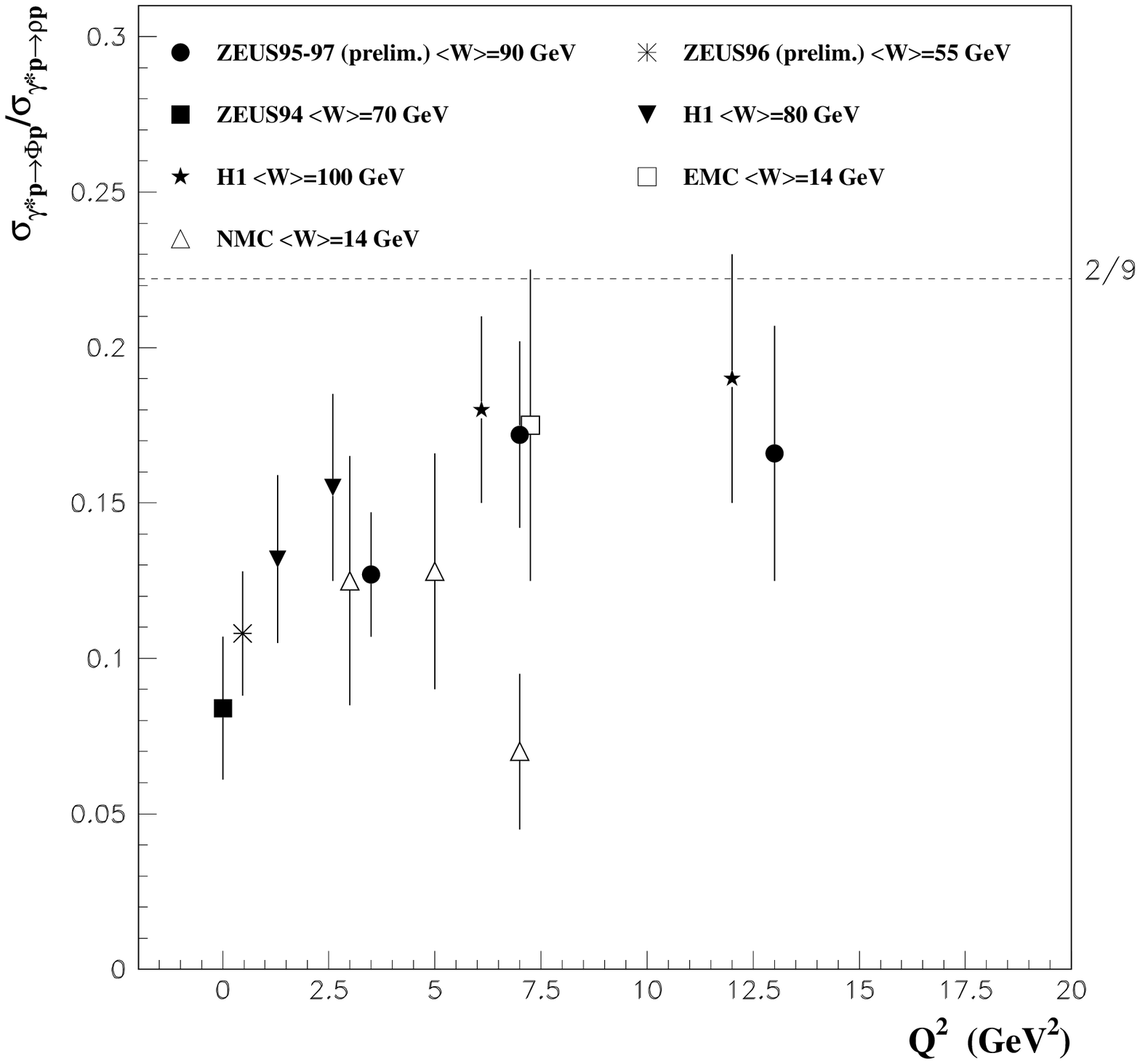,%
            width=6.5cm,height=6.0cm}}
   \put(6.2,-0.2){\epsfig{file=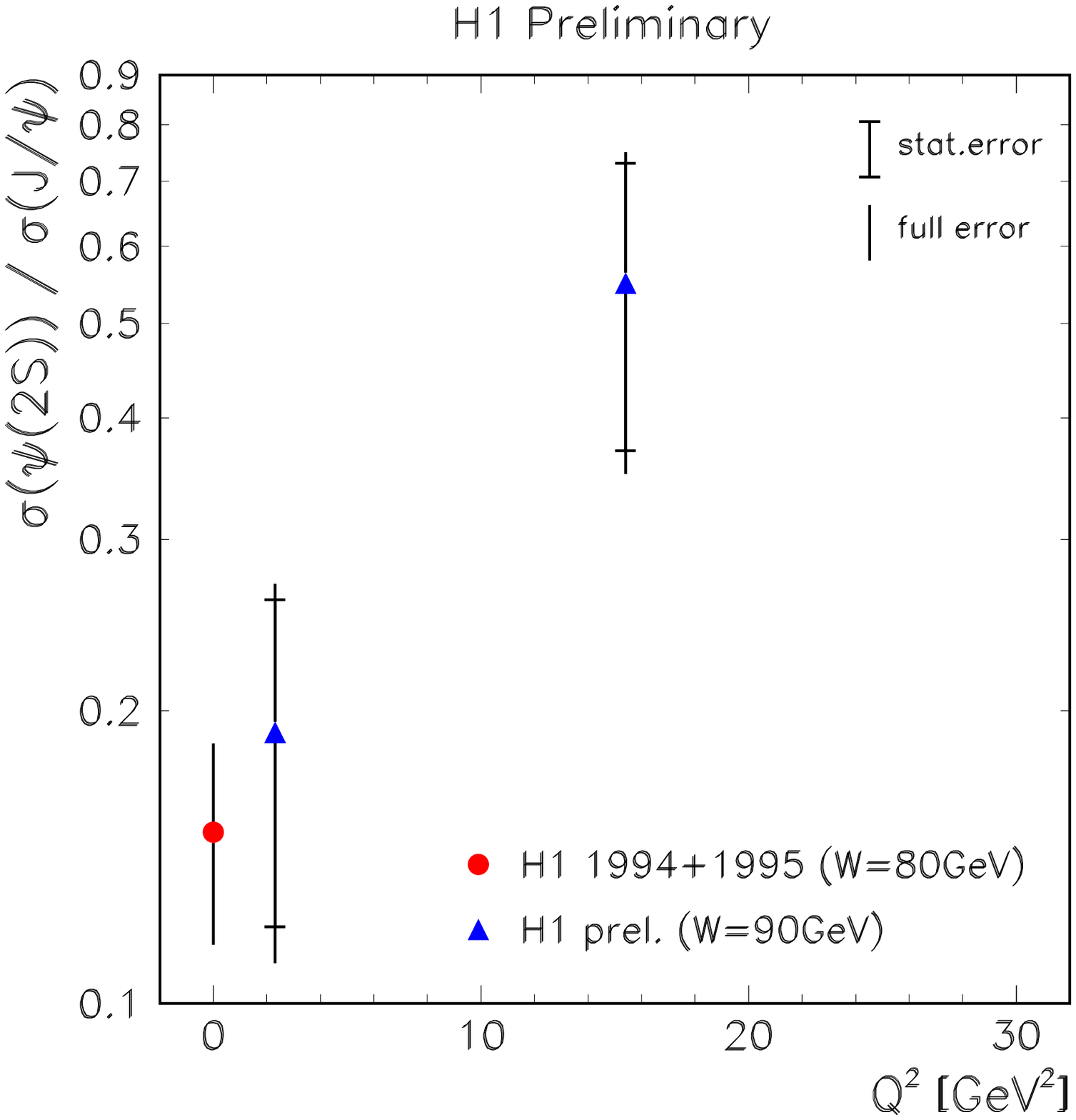,%
            width=6.5cm,height=6.0cm}}
  \end{picture}
\vspace{-0.7cm}
\caption{Ratio $R$ of cross sections for a) $\phi$ and \rh ;
b) $\psi(2s)$ and \jpsi\ meson production, as a function of \qsq.}
\label{fig:VM_ratio}
\end{figure}
\vspace{-0.7cm}
\end{center}

The case of the $\psi(2s) / \psi$ ratio illustrates the interesting 
phenomenon of the ``scanning'' of the VM wave function as \qsq\ varies.
Because of the node in the $\psi(2s)$ wave function, which induces 
approximately cancelling contributions in the production amplitude, the 
photoproduction of $\psi(2s)$ mesons is small.
As \qsq\ increases, the transverse size of the $q \bar q$ pair decreases, 
thus avoiding the cancellation effect.
The resulting increase with \qsq\ of the cross section ratio is illustrated
in Fig.~\ref{fig:VM_ratio}b~\cite{h_jpsi_DIS}, the asymptotic limit being 
computed to be of the order of 0.5~\cite{kopel,fks}.

\subsection{Mass Distribution}

\begin{center}
\begin{figure}[htbp] \unitlength 1.0 cm
\vspace{-0.8cm}
  \begin{picture}(12.0,6.0)
   \put(0.0,-0.2){\epsfig{file=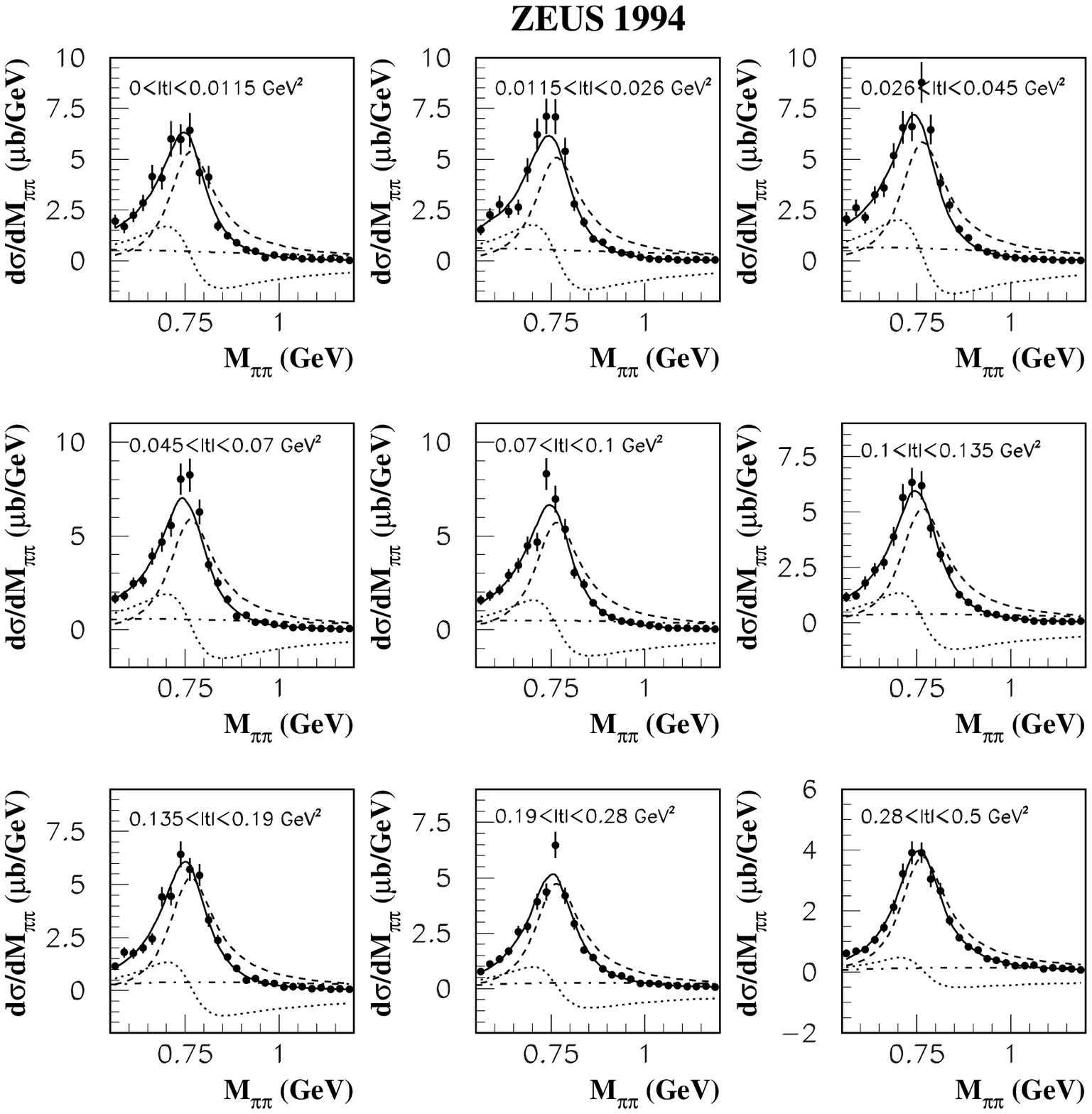,%
            width=7.0cm,height=6.0cm}}
   \put(6.3,-0.2){\epsfig{file=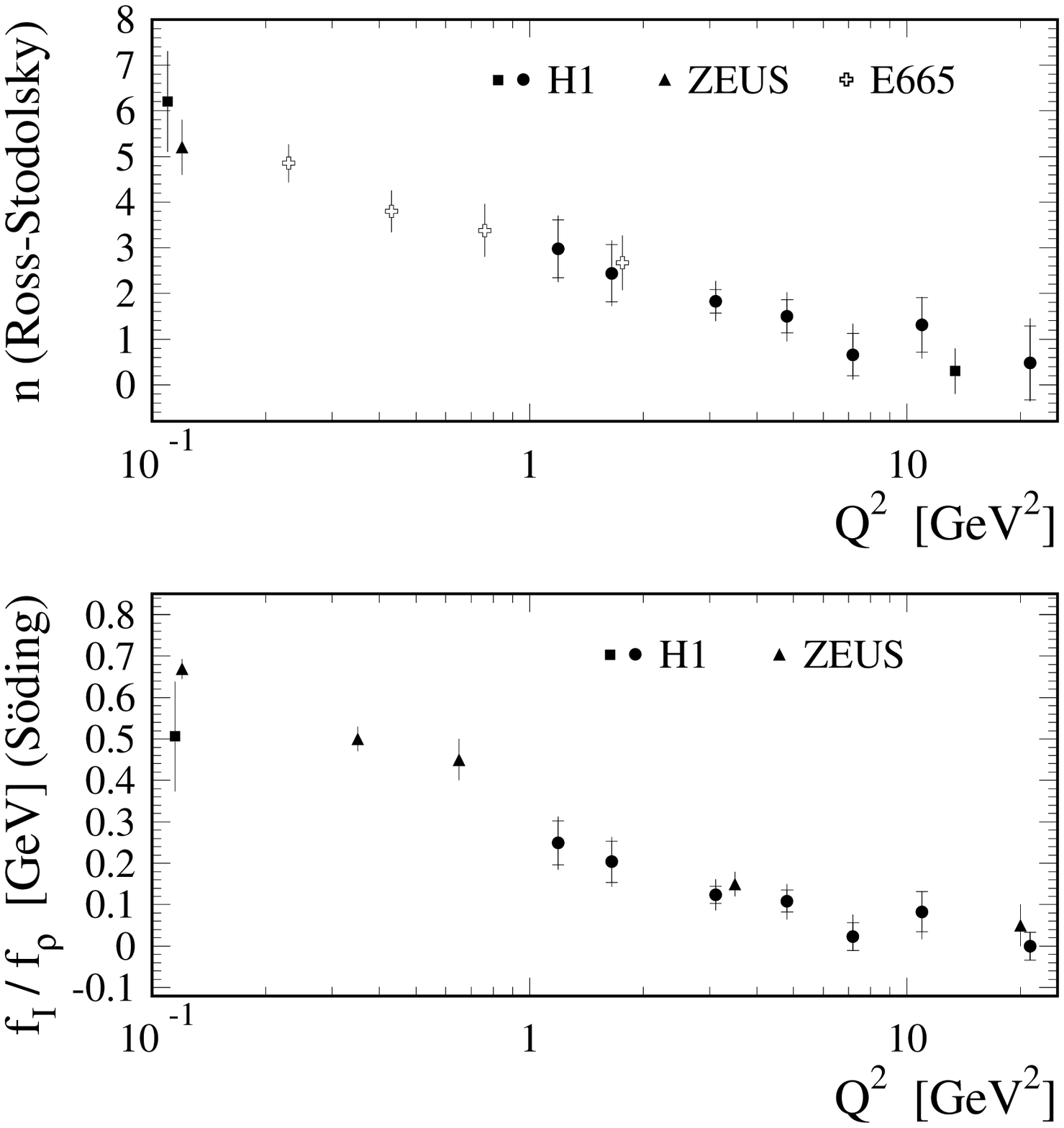,%
            width=5.7cm,height=5.6cm}}
  \end{picture}
\vspace{-0.5cm}
\caption{a) Mass distribution $M_{\pi \pi}$ for \rh\ photoproduction,
for different values of \modt~\protect\cite{zeus_phottt}; the dashed curves
represent the Breit-Wigner distribution, the dotted curves the interference
with non-resonant pion pair, and the full curves the sum. 
b) Two parameterisations~\protect\cite{soding,ross_sto} of the skewing of 
the $M_{\pi \pi}$ mass distribution for \rh\ production, as a function of 
\qsq~\protect\cite{E665,zeus_phot,zeus_phottt,h_rho_phot,z_rho_jpsi_DIS,h_rho_DIS}.}
\label{fig:skewing}
\vspace{-1.0cm}
\end{figure}
\end{center}

For \rh\ photoproduction, the ($\pi,\pi$) mass distribution is 
distorted with respect to a (relativistic) Breit-Wigner distribution,
with an excess of events at small masses and a deficit at large masses.
This phenomenon, known as ``skewing'', is attributed to the interference
between resonant \rh\ production and non-resonant pion pair production, the
interference changing sign at the resonance pole~\cite{soding}.
The skewing is observed to decrease in photoproduction as \modt\ 
increases~\cite{zeus_phottt} (see Fig.~\ref{fig:skewing}a).
The skewing also decreases with increasing 
\qsq\ as seen in Fig.~\ref{fig:skewing}b for two different 
parameterisations~\cite{soding,ross_sto}.

%

\section{Conclusions}
%
Abundant data have been collected at HERA on diffractive production of
light and heavy vector mesons, in the presence of the scales $Q^2$, $m_q$ 
and $t$.

A strong energy dependence of the cross section is observed for \jpsi\ 
production; an indication is found for a similar behaviour for \rh\ 
mesons at high \qsq.
In the light of perturbative QCD, with the pomeron modelled as a gluon pair,
these features are interpreted as reflecting the strong increases of the 
gluon distribution in the proton at high energy, and quantitative agreement is
reached for \jpsi\ production.
The \qsq\ dependence of VM production is also qualitatively explained in pQCD
approaches.

The ratio of the longitudinal to transverse photon cross sections for \rh\
production increases rapidly with \qsq, but this increase is non-linear for
$Q^2 \gsim 2$ \gevsq.
This behaviour has been reproduced recently by a model based on QCD.
More generally, the full set of \rh\ meson spin density matrix elements has 
been measured.
The correct hierarchy between scattering amplitudes and the magnitude 
of the dominant helicity-flip amplitude are also qualitatively reproduced
in a QCD approach. 

In summary, great progress has been made in the understanding of VM
production at high energy when a hard scale is present ($m_c$, \qsq). 
This contributes significantly to the understanding of 
diffraction in a QCD framework.

\section*{Acknowledgements}
It is a pleasure to thank the organisors for a pleasant and fruitful 
Symposium, and my colleagues in H1 and ZEUS, in particular B. Clerbaux and 
P. Newmann, for numerous interesting discussions on diffraction.

\end{document}